\documentclass[11pt]{article}
\usepackage{bbm}
\usepackage{amsfonts}
\usepackage{mathrsfs}
\usepackage{pifont}
\usepackage{amsmath}
\usepackage{amssymb}
\usepackage{graphics}
\begin{document}
\title{ One-way Unlocalizable Information Deficit}

\author{ Xue-Na Zhu$^{1}$
 \ \ \ Shao-Ming Fei$^{2,3}$ \\
{\footnotesize  {$^1$Department of Mathematics,
 South China University of Technology, Guangzhou
510640, P.R.China}} \\
{\footnotesize{
  $^2$School of Mathematical Sciences, Capital Normal University,
Beijing 100048, China}}\\
{\footnotesize{$^3$Max-Planck-Institute for Mathematics in the Sciences, 04103
Leipzig, Germany}}}

\date{}

\maketitle

{\bf Abstract:} We introduce one-way unlocalizable information deficit
with respect to the one-way information deficit, similar to the definition of
one-way unlocalizable quantum discord with respect to one-way quantum discord.
The properties of the one-way unlocalizable information deficit
and the relations among one-way unlocalizable  information deficit,
one-way unlocalizable quantum discord, one-way quantum discord,
one-way information deficit and other quantum correlations are investigated.
Analytical formulas of the one-way unlocalizable quantum discord are given to detailed examples.

\section*{1. Introduction}
Quantum entanglement \cite{qe1} is of special importance in quantum information processing
such as quantum teleportation, dense coding and remote state preparation.
To quantify quantum entanglement various entanglement measures have been
suggested \cite{ASHD}. Considerable efforts have been made to estimate these entanglement measures and to
investigate the roles played by these measures in different information processing.
The one-way unlocalizable entanglement in terms of entanglement of assistance has been
presented in Ref.\cite{FGJ}. It is shown
that the polygamous nature of distributed quantum entanglement in multipartite systems is strongly
related to this unlocalizable character.

Quantum correlations other than the quantum entanglement have been also extensively explored recently.
It has been shown that some quantum information processing could be carried out without quantum entanglement.
For instance, the quantum discord \cite{qd} plays an important role in some quantum information processing
like assisted optimal state discrimination, in which only one side discord is required in the optimal
process of assisted state discrimination,
while another side discord and entanglement is not necessary \cite{ost}.

With respect to the one-way quantum discord,
one-way unlocalizable quantum discord has been introduced and studied in \cite{ZHY}.
In this paper, with respect to the one-way information deficit,
we introduce and study the one-way unlocalizable information deficit.
Systematic relations among one-way unlocalizable information deficit,
one-way unlocalizable quantum discord, one-way quantum discord,
one-way information deficit and other quantum correlations are presented.
Trade off relations are discussed.

\section*{2. One-way unlocalizable  information deficit}

\subsection*{2.1.  Definition}
Quantum correlation is emerging as a primitive notion in physics
following an essential extension of classical Shannon
information theory into the quantum domain. There have been many
different definitions of measures for quantum correlations.

Let $H_{A}$ and $H_{B}$ be the $m$ and $n$-dimensional
$(m \leq n)$ vector spaces, respectively.
The quantum discord of a bipartite quantum state $\rho^{AB}\in H_{A}\otimes H_{B}$
is defined by \cite{wha,howh},
\begin{equation}\label{D}
\delta^{\leftharpoonup}(\rho^{AB})=S(\rho^{B})-S(\rho^{AB})+\min_{\{\Pi_i^B\}}\sum_{i}p_iS(\rho^{A}_i),
\end{equation}
where $S(\rho)=-Tr[\rho\log_2\rho]$ is the von Neumann entropy,
$\rho^{B}=Tr_{A}(\rho^{AB})$ is the reduced density matrix of system $B$,
$p_i=tr(I_A\otimes \Pi_i)\rho^{AB} (I_A\otimes \Pi_i)$ with $I_A$ the identity operator on subsystem $A$,
$\rho^{A}_i=tr_B(I_A\otimes \Pi^{B}_i\rho^{AB})/p_i$ is the state of subsystem $A$ after the measurement on $B$,
$\Pi_i^B=|i\rangle\langle i|$ is the von Neumann measurement on $B$ satisfying
$\sum_i\Pi_i^B=I_B$, with $I_B$ the identity operator on $B$, $|i\rangle$, $i=1,...n$,
is the computational basis.

Inspired by the definition of unlocalizable entanglement in \cite{FGJ}, the authors
in \cite{ZHY} provided the quantity one-way unlocalizable quantum discord,
\begin{equation}\label{UD}
\delta_{\mu}^{\leftharpoonup}(\rho^{AB})=S(\rho^{B})-S(\rho^{AB})+\max_{\{\Pi^{B}_i\}}\sum_{i}p_iS(\rho^{A}_i).
\end{equation}

Closely related to the one-way quantum discord,
the one-way information deficit \cite{MPRJAUB,AHD} is defined as the
minimal increase of entropy after a von Neumann measurement on $B$ \cite{AHD}:
\begin{equation}\label{informationdeficit}
\Delta^{\leftharpoonup}(\rho^{AB})=\min_{\{\Pi^B_{i}\}}S(\sum_{i}\Pi^B_{i}\rho^{AB}\Pi^B_{i})-S(\rho^{AB}).
\end{equation}

Similar to the quantum discord (\ref{D}) and the one-way unlocalizable quantum discord (\ref{UD}),
corresponding to the one-way information deficit (\ref{informationdeficit}), we define
the one-way unlocalizable  information deficit,
\begin{equation}\label{minformationdeficit}
\Delta_{\mu}^{\leftharpoonup}(\rho^{AB})=\max_{\{\Pi^B_{i}\}}S(\sum_{i}\Pi^B_{i}\rho^{AB}\Pi^B_{i})-S(\rho^{AB}).
\end{equation}

Dual to the one-way information deficit $\Delta^{\leftharpoonup}(\rho^{AB})$,
$\Delta_{\mu}^{\leftharpoonup}(\rho^{AB})$
is the maximum distance of relative entropy from the state $\rho^{AB}$ to the set
$S^{\leftharpoonup}$ which can be created reversibly under one-way communications.

If the maximum in (\ref{minformationdeficit}) is taken over all the von Neumann measurements $\{\Pi^B_{i}\}$
which do not disturb reduced states $\rho^{B}=Tr_{A}(\rho^{AB})$ locally,
then the one-way unlocalizable information deficit $\Delta_{\mu}^{\leftharpoonup}(\rho^{AB})$
is just the relative entropy of nonlocality \cite{042325}:
\begin{equation}
\textrm{N}^\leftarrow_{RE}(\rho^{AB})=\max_{\{\Pi^B_i\}}[S(\tilde{\rho}^{AB})-S(\rho^{AB})],
\end{equation}
where $\tilde{\rho}^{AB}=\sum_{i}(I_A\otimes \Pi_i^B)\rho^{AB}(I_A\otimes \Pi_i^B)$.
Hence the one-way unlocalizable information deficit $\Delta_{\mu}^{\leftharpoonup}(\rho^{AB})$
is equal to the relative entropy of nonlocality $\textrm{N}^\leftarrow_{RE}(\rho^{AB})$
for a set of special states $\rho^{AB}$ such that
$\rho^{B}$ is invariant under all von Neumann measurements $\{\Pi^B_{i}\}$.

\subsection*{2.2. Polygamy inequality }

Similar to the relations between $\delta^{\leftharpoonup}(\rho^{AB})$
and $\delta_{\mu}^{\leftharpoonup}(\rho^{AB})$ investigated in \cite{FGJ},
in the following we study the relations between $\Delta^{\leftharpoonup}(\rho^{AB})$ and
$\Delta_{\mu}^{\leftharpoonup}(\rho^{AB})$, the relations among
the quantum discord, the one-way unlocalizable quantum discord, the one-way information deficit and
the one-way unlocalizable information deficit, as well as tradeoff relations.

We first present some properties of the one-way unlocalizable information deficit.
From Refs.\cite{MR,MJ}, any partial von Neumann measurement on $B$ can be modeled
as an indirect measurement with an apparatus $Q$ initialized in a fixed pure state $|0\rangle^{M}$,
$\rho_1=|0\rangle^{M}\langle0|\otimes\rho_{AB}$, and applying a unitary operator $U$ on the whole
state, $\rho_2=U\rho_1U^{\dag}$, where $U= I_A\otimes U_{MB}$ and
$Tr_M[U(|0\rangle^{M}\langle0|\otimes\rho_{AB}U^{\dag}]=\sum_{i}\Pi^B_{i}\rho^{AB}\Pi^B_{i}$.
As an important measure of
quantum entanglement, the distillable entanglement $E_D(\rho^{AB})$ defined in \cite{CDJ,MBS}
is the asymptotic number of standard singlets
that can be prepared from a system in state $\rho^{AB}$ by local
operations. From \cite{AHD}, one has
$E_{D}^{M|AB}(U\rho_1U^{\dag})=S(\sum_{i}\Pi^B_{i}\rho^{AB}\Pi^B_{i})-S(\rho^{AB})$.
Hence if a bipartite state $\rho^{AB}$ has nonzero quantum
discord $\delta^{\leftarrow}>0$, any von Neumann measurement
on $B$ creates distillable entanglement between the measurement
apparatus and the total system $AB$. The maximal
distillable entanglement created in a von Neumann measurement
on $B$ is equal to the one-way unlocalizable information deficit:
$\Delta^{\leftharpoonup}_{\mu}(\rho^{AB})=\max_{U}E_{D}^{M|AB}(U\rho_1U^{\dag})$.
Therefore, the one-way unlocalizable information deficit has the properties:

i) $\Delta^{\leftharpoonup}_{\mu}(\rho^{AB})$ does not increase under arbitrary quantum operations $\Lambda_{B}$
on B,
\begin{equation}\label{notin}
\Delta^{\leftharpoonup}_{\mu}(\Lambda_B(\rho^{AB}))\leq\Delta^{\leftharpoonup}_{\mu}(\rho^{AB}),
\end{equation}
as $E_{D}^{M|AB}$ does not increase under local operations and classical communications.

ii) $\Delta^{\leftharpoonup}_{\mu}(\rho^{AB})$ does not increase on average under stochastic
local operations and classical communications (SLOCC):
\begin{equation}\label{inav}
\sum_iq_i\Delta^{\leftharpoonup}_{\mu}(\sigma_i^{AB})\leq\Delta^{\leftharpoonup}_{\mu}(\rho^{AB}),
\end{equation}
where $q_i=Tr[V_i\rho^{AB} V_i^{\dag}]$, $\sigma_i=V_i\rho^{AB} V_i^{\dag}/q_i$, and ${V_i}$ are Kraus operators
characterizing a local quantum operation on $B$ with $\sum_iV_i^{\dag}V_i=I$.
Inequality (\ref{inav}) is due to that the distillable entanglement does not increase on average under SLOCC: $\sum_iq_iE(\sigma_i^{AB})\leq E(\rho^{AB})$.

From the definitions  (\ref{D}), (\ref{UD}), (\ref{informationdeficit}) and (\ref{minformationdeficit}), we
have the following lower bounds of the one-way unlocalizable information deficit.

{\bf Theorem 1:} For any bipartite quantum state $\rho^{AB}$, we have
\begin{equation}\label{ineq}
\begin{aligned}
&\Delta^{\leftharpoonup}_{\mu}(\rho^{AB})\geq\Delta^{\leftharpoonup}(\rho^{AB});\\[1mm]
&\Delta^{\leftharpoonup}_{\mu}(\rho^{AB})\geq\delta^{\leftharpoonup}_{\mu}(\rho^{AB});\\[1mm]
&\Delta^{\leftharpoonup}_{\mu}(\rho^{B})\geq\Delta^{\leftharpoonup}(\rho^{AB})-\delta^{\leftharpoonup}(\rho^{AB}).
\end{aligned}
\end{equation}

{\sf [Proof]}: The first inequality is easily obtained from the
definitions of $\Delta^{\leftharpoonup}_{\mu}(\rho^{AB})$ and $\Delta^{\leftharpoonup}(\rho^{AB})$.

For $p_i=Tr[\Pi^B_{i}\rho^{AB}\Pi^B_{i}]$ and $\rho^{A}_i=\Pi^B_{i}\rho^{AB}\Pi^B_{i}/p_i$, one has the following equality \cite{AHD}:
\begin{equation}\label{deng}
\sum_ip_iS(\rho^{A}_i)=S(\sum_{i}\Pi^B_{i}\rho^{AB}\Pi^B_{i})-S(\sum_{i}\Pi^B_{i}\rho^{B}\Pi^B_{i}).
\end{equation}
According to the definition of the one-way unlocalizable quantum discord, we obtain
\begin{equation}\label{ineq}
\begin{aligned}
\delta^{\leftharpoonup}_{\mu}(\rho^{AB})&=
S(\rho^{B})-S(\rho^{AB})+
\max_{\{\Pi^B_{i}\}}\{S(\sum_{i}\Pi^B_{i}\rho^{AB}\Pi^B_{i})-S(\sum_{i}\Pi^B_{i}\rho^{B}\Pi^B_{i})\}\\[3mm]
&\leq
S(\rho^{B})-S(\rho^{AB})+\max_{\{\Pi^B_{i}\}}S(\sum_{i}\Pi^B_{i}\rho^{AB}\Pi^B_{i})
-\min_{\{\Pi^B_{i}\}}S(\sum_{i}\Pi^B_{i}\rho^{B}\Pi^B_{i})\\[3mm]
&=\Delta^{\leftharpoonup}_{\mu}(\rho^{AB})-\Delta^{\leftharpoonup}(\rho^{B})
=\Delta^{\leftharpoonup}_{\mu}(\rho^{AB}).\\[3mm]
\end{aligned}
\end{equation}
The last inequality in (\ref{ineq}) can be proved similarly,
\begin{equation}
\begin{aligned}
\delta^{\leftharpoonup}(\rho^{AB})&=S(\rho^{B})-S(\rho^{AB})+
\min_{\{\Pi^B_{i}\}}\{S(\sum_{i}\Pi^B_{i}\rho^{AB}\Pi^B_{i})-S(\sum_{i}\Pi^B_{i}\rho^{B}\Pi^B_{i})\}\\[3mm]
&\geq\displaystyle
S(\rho^{B})-S(\rho^{AB})+\min_{\{\Pi^B_{i}\}}S(\sum_{i}\Pi^B_{i}\rho^{AB}\Pi^B_{i})-\displaystyle
\max_{\{\Pi^B_{i}\}}S(\sum_{i}\Pi^B_{i}\rho^{B}\Pi^B_{i})\\[3mm]
&=\Delta^{\leftharpoonup}(\rho^{AB})-\Delta^{\leftharpoonup}_{\mu}(\rho^{B}).
\end{aligned}
\end{equation}
Here, in fact, $\Delta^{\leftharpoonup}_{\mu}(\rho^{B})=\log_{2}n-S(\rho^B)$.
$\hfill\Box$

The relationship between $\delta^{\leftharpoonup}$ (defined on single copies)
and $\Delta^{\leftharpoonup}$ was shown in \cite{whz}. The one-way information deficit is non-negative
and zero only for states with zero quantum discord.
To compare the one-way unlocalizable deficit with other measures of the
quantum correlation: the one-way unlocalizable quantum discord,
we consider the Bell-diagonal states,
\begin{equation}\label{mm}
\rho^{AB}_m=\frac{1}{4}(I^{A}\otimes I^{B}+\sum_{i=1}^{3}c_i\sigma_i\otimes \sigma_i),
\end{equation}
where $c_i$ are real numbers and $\sigma_i$ are pauli matrices.

For the state $\rho^{AB}_m$, one has
$\Delta^{\leftharpoonup}(\rho_m^{AB})=\delta^{\leftharpoonup}(\rho_m^{AB})$.
It can be also proven that
$\Delta^{\leftharpoonup}_{\mu}(\rho_m^{AB})=\delta^{\leftharpoonup}_\mu(\rho_m^{AB})$.
A von Neumann measurement $\{\Pi_0^B, \Pi_1^B\}$ of a two-qubit system
can be characterized by a unit vector $n=(n_1,n_2,n_3)^{T}$ on the Bloch sphere \cite{LJXW},
$$
\begin{array}{lc}\displaystyle
\Pi_0^B=\frac{1}{2}(I^B+\sum_{i=1}^{3}n_i \sigma_i^B),~~~~~
\Pi_1^B=\frac{1}{2}(I^B-\sum_{i=1}^{3}n_i \sigma_i^B).
\end{array}
$$
From the Eq. (\ref{mm}), we have $\rho^B_m=Tr_A(\rho^{AB}_m)={I^B}/{2}$.
Set $n_1=\cos({x}/{2})\sin({y}/{2})$, $n_2=\cos({x}/{2})\cos({y}/{2})$
and $n_3=\sin({x}/{2})$.
We have $S(\sum_{i}\Pi_i^B\rho^B_m\Pi_i^B)=S({I^B}/{2})=S(\rho^B_m)$. From Eq.(\ref{minformationdeficit}),
we obtain $\Delta^{\leftharpoonup}_{\mu}(\rho^{AB}_m)=\delta^{\leftharpoonup}_\mu(\rho^{AB}_m)$.

Moreover, from \cite {042325} we have
\begin{equation}
\Delta^{\leftarrow}_{\mu}(\rho_m^{AB})=\delta^{\leftarrow}_{\mu}(\rho_m^{AB})=f(c_{\min})-f(c_1,c_2,c_3),
\end{equation}
where
$$
f(x)=-\frac{1+x}{2}\log_2(\frac{1+x}{2})-\frac{1-x}{2}\log_2(\frac{1-x}{2}),
$$
$c_{\min}=\min\{|c_1|,|c_2|,|c_3|\}$ and
$$
\begin{array}{rcl}
f(c_1,c_2,c_3)&=&
\displaystyle -(\frac{1-c_1-c_2-c_3}{4}\log_2(\frac{1-c_1-c_2-c_3}{4})\\[3mm]
&&+\displaystyle\frac{1-c_1+c_2+c_3}{4}\log_2(\frac{1-c_1+c_2+c_3}{4})\\[3mm]
&&+\displaystyle\frac{1+c_1-c_2+c_3}{4}\log_2(\frac{1+c_1-c_2+c_3}{4})\\[3mm]
&&+\displaystyle\frac{1+c_1+c_2-c_3}{4}\log_2(\frac{1+c_1+c_2-c_3}{4}))-1.
\end{array}
$$

Generally one has the following conclusion:

{\bf Theorem 2}~
For any bipartite quantum state $\rho^{AB}$, the one-way unlocalizable information
deficit is zero if and only if the one-way unlocalizable quantum discord is zero.

{\sf [Proof]}~
If the one-way unlocalizable information
deficit is zero: $\Delta^{\leftharpoonup}_{\mu}(\rho^{AB})=0$, from the first inequality of Theorem 1,
we obtain that $\Delta^{\leftharpoonup}(\rho^{AB})=0$.  Then from Eqs.(\ref{informationdeficit}) and (\ref{minformationdeficit}), we have
 $\max_{\{\Pi^B_{i}\}}S(\sum_{i}\Pi^B_{i}\rho^{AB}\Pi^B_{i})=S(\rho^{AB})$
and
$\min_{\{\Pi^B_{i}\}}S(\sum_{i}\Pi^B_{i}\rho^{AB}\Pi^B_{i})=S(\rho^{AB})$.
Therefore for all $\{\Pi_i^B\}$, formula $S(\sum_{i}\Pi^B_{i}\rho^{AB}\Pi^B_{i})=S(\rho^{AB})$ is true.
By using Eq.(\ref{deng}), one proves the result.

If the one-way unlocalizable quantum discord is zero: $\delta^{\leftharpoonup}_{\mu}(\rho^{AB})=0$, since
$0\leq\delta^{\leftharpoonup}(\rho^{AB})\leq\delta^{\leftharpoonup}_{\mu}(\rho^{AB})$, we have
$\delta^{\leftharpoonup}(\rho^{AB})=0$. Hence $\rho^{AB}$ has the form
$\rho^{AB}=\sum_ip_i\rho_i^A\otimes|i\rangle\langle i|$, and $\delta^{\leftharpoonup}_{\mu}(\rho^{AB})=0$.
$\hfill\Box$

\subsection*{2.3. Discussions on tradeoff relations }

In the following we study the tradeoff relations.
The one-way unlocalizable quantum entanglement is defined by \cite{ZHY}:
\begin{equation}
\emph{S}^{\leftarrow}_{\chi}(\rho^{AB})=\min_{\{\Pi^B_i\}}[S(\rho^A)-\sum_ip_iS(\rho^A_i)].
\end{equation}
For a tripartite pure state $|\psi\rangle^{ABC}$, in terms of the Buscemi Gour-Kim equation \cite{ZHY} a
tradeoff relation between the one-way unlocalizable quantum discord for $A$ and $B$ with the measurement on $B$
and the one-way unlocalizable quantum entanglement for $C$ and $B$ with the measurement on $B$
has been obtained in \cite{ZHY},
\begin{equation}\label{dD}
\delta^{\leftharpoonup}_{\mu}(\rho^{AB})=S(\rho^B)- \emph{S}^{\leftarrow}_{\chi}(\rho^{BC}).
\end{equation}

For states that all the measurements $\{\Pi_i^{B}\}$ do not disturb $\rho^{B}$ locally,
there is a tradeoff relation between the one-way unlocalizable
deficit for $A$ and $B$ with the measurement on $B$
and the one-way unlocalizable quantum entanglement for $C$ and $B$ \cite{042325},
\begin{equation}\label{Dd}
\Delta^{\leftharpoonup}_{\mu}(\rho^{AB})=S(\rho^B)- \emph{S}^{\leftarrow}_{\chi}(\rho^{BC}).
\end{equation}

According to the tradeoff relations (\ref{dD}) and (\ref{Dd}), if $\rho^{AB}$ satisfies the particular
condition that all the von Neumann measurements $\{\Pi^i_B\}$ do not disturb $\rho^B$ locally,
for instance, the state $\rho^{AB}_m$ in (\ref{mm}), we have
\begin{equation}\label{Ddd}
\begin{array}{lc}
\displaystyle\Delta^{\leftharpoonup}_{\mu}(\rho^{AB})=\delta^{\leftharpoonup}_{\mu}(\rho^{AB}).
\end{array}
\end{equation}

\section*{3. Conclusion }
We have introduced the one-way unlocalizable  information deficit.
The essential properties of the One-way unlocalizable  information deficit
and some foundational relations among the one-way unlocalizable  information deficit,
the one-way unlocalizable quantum discord, the one-way quantum discord,
the one-way information deficit and other quantum correlations have been investigated.
As different kinds of quantum correlations play different roles in different quantum information
processing, these properties and relations might allow to highlight the deep
relations among information processing and correlations.

\end{document}